\documentstyle[12pt]{article}
\def\be{\begin{equation}}
\def\ee{\end{equation}}
\def\bea{\begin{eqnarray}}
\def\eea{\end{eqnarray}}
\def\kt{{k_{\perp}}}
\def\as{\alpha_S}

\def\ga{\gamma}
\def\gan{\gamma_N}
\def\G{{\Gamma}}

\def\ms{$\overline{{\rm MS}}$}
\def\cf{{\cal F}}

\def\np#1#2#3{Nucl.\ Phys.\ B#1 (19#3) #2}
\def\pl#1#2#3{Phys.\ Lett.\ #1B (19#3) #2}
\def\pr#1#2#3{Phys.\ Rev.\ D #1 (19#3) #2}

\def\sj#1#2#3{Sov.\ J.\ Nucl.\ Phys.\ #1 (19#3) #2}
\def\zp#1#2#3{Z.\ Phys.\ C#1 (19#3) #2}

\begin{document}
\begin{titlepage}
\renewcommand{\thefootnote}{\fnsymbol{footnote}}
\begin{flushright}
     hep-ph/9608310 \\ DFF 254-7-96 \\   July 1996
     \end{flushright}
\par \vspace{10mm}
\begin{center}
{\Large \bf $k_{\perp}$-factorization and perturbative invariants at \\
small $x$}
\end{center}
\par \vspace{2mm}
\begin{center}
{\bf Stefano Catani}~\footnote{Talk given at International
Workshop on Deep Inelastic Scattering and Related Phenomena, {\it DIS 96},
Rome, Italy, 15-19 April 1996. 
A shorter version will appear in the Proceedings.} \\

\vspace{5mm}

{I.N.F.N., Sezione di Firenze}\\
{and Dipartimento
di Fisica, Universit\`a di Firenze}\\
{Largo E. Fermi 2, I-50125 Florence, Italy}

\end{center}

\par \vspace{2mm}
\begin{center} {\large \bf Abstract} \end{center}
\begin{quote}
I review the basic idea of $k_{\perp}$-factorization and its 
relation to collinear factorization. Theoretical results in resummed 
perturbation theory are summarized and the example of the heavy-flavour 
structure functions is explicitly considered. Using these results one can 
investigate the small-$x$ behaviour of quantities that are independent of the
non-perturbative parton densities. In particular, one can introduce physical
anomalous dimensions that relate the scaling violations in different hadronic
observables.
\end{quote}
\vspace{1cm}
\begin{flushleft}
    DFF 254-7-96 \\   July 1996
\end{flushleft}

\end{titlepage}

\section{$k_{\perp}$-factorization in hadron collisions at high energy}
\label{ktsec}

Hadron collisions at large transferred momentum $Q$ ($Q \gg \Lambda$, $\Lambda$
being the QCD scale) can be studied in QCD perturbation theory by computing
the corresponding cross sections as power-series expansions in $\as(Q^2)$. In
the high-energy regime ${\sqrt S} \gg Q$ (${\sqrt S}$ the centre-of-mass energy)
or, equivalently, at small values of the ratio $x=Q^2/S$, the coefficients
of these power-series expansions contain logarithmically-enhanced contributions 
of the type $\ln^n x$. As soon as $\as \ln 1/x \sim 1$, the fixed-order 
expansion in $\as$ is no longer reliable. The higher-order contributions  
$(\as \ln x)^n$ have to be evaluated and, possibly, resummed to all orders in 
perturbation theory.

The basic theoretical input for the resummation is provided by the BFKL 
equation~\cite{BFKL} for the gluon distribution $\cf(x,\kt;Q_0)$. 
This distribution describes 
the evolution of an initial-state gluon with momentum $p^\mu$ ($p^2=Q_0^2 \sim
1 {\rm GeV}^2$) into an off-shell gluon with momentum $k^\mu = xp^\mu + \kt^\mu$
($- k^2 = - \kt^2 \gg Q_0^2$). The evolution process is obtained by radiating 
final-state partons with momenta $\kt_i$ ($\kt = - \sum_i \kt_i$). The BFKL 
equation thus resums $(\as \ln x)^n$ terms due to gluon evolution over the
large rapidity gap $y=\ln 1/x$. Since these terms are produced by emission of 
partons with any transverse momentum $\kt_i$, no $\kt$-ordering is embodied in
the BFKL equation.

Gluons are not directly observable in scattering processes. Having at our 
disposal 
the BFKL equation, we still have to relate it to physical cross sections.
The relation is provided by the $\kt$-factorization theorem~\cite{CCH,CE,LRSS}. 
In the case of processes involving a single incoming hadron, like, for instance,
in deep-inelastic lepton-hadron scattering (DIS), the cross section is written 
as follows
\bea\label{ktfor}
\sigma(x,Q^2) &=& \int_x^1 \frac{dz}{z} \int d^2\kt \;{\hat \sigma}(x/z,Q^2;\kt) 
\;\cf(z,\kt) \;\;,\\
\label{ktfg}
\cf(x,\kt) &=& \int_x^1 \frac{dz}{z} \;\cf(x/z,\kt;Q_0) \;{\bar f}_g(z,Q_0) 
\;\;.
\eea
In Eq.~(\ref{ktfg}) the lower-end gluon $(p^2=Q_0^2)$ of the BFKL distribution
$\cf(x,\kt;Q_0)$ is first coupled to the 
incoming hadron  via a non-perturbative (but process-independent) gluon density
${\bar f}_g(x,Q_0)$. Then, in Eq.~(\ref{ktfor}) the upper-end gluon $k$ 
is coupled to the scattering process via the partonic cross section 
${\hat \sigma}$. Note that, since the $(\as \ln x)^n$ contributions to 
$\sigma(x,Q^2)$ are due to the emission of partons with any $\kt_i$, in order 
not
to spoil the consistency of small-$x$ resummation, the partonic cross section
${\hat \sigma}(z,Q^2;\kt)$ has to be properly defined (and perturbatively 
computed) in terms
of an incoming off-shell $(k^2=\kt^2)$ gluon. Accordingly, an unconstrained 
$\kt$-integration is involved in the factorization formula (\ref{ktfor}).
 
\section{Collinear factorization and small-$x$ resummation}
\label{colsec}

Equations (\ref{ktfor},\ref{ktfg}) have to be compared with the customary
perturbative QCD formulae as obtained by applying the factorization theorem of
collinear singularities~\cite{CSS}. According to the latter the hadronic cross 
section is written in the following way
\be\label{colfor}
\sigma(x,Q^2) = \sigma_0(Q^2) \sum_a \int_x^1 \frac{dz}{z} \;C^a(x/z,\as(Q^2)) 
\;{\tilde f}_a(z,Q^2) \;\;,
\ee
where the rescaling factor $\sigma_0(Q^2)$ has been introduced to make 
dimensionless the coefficient functions $C^a(x,\as(Q^2))$. The parton densities
${\tilde f}_a(x,Q^2)$ fulfil the Altarelli-Parisi equations~\cite{AP}:
\be\label{eveq}
\frac{d {\tilde f}_a(x,Q^2)}{d \ln Q^2} = \sum_b \int_x^1 dz \;
P_{ab}(z,\as(Q^2)) \;{\tilde f}_b(x/z,Q^2) \;\;.
\ee
Both the coefficient functions $C^a(x,\as)$ and the Altarelli-Parisi
splitting functions $P_{ab}(x,\as)$ are computable in perturbation theory order 
by order in $\as$.
 
The main difference between the factorization formulae (\ref{ktfor}) and 
(\ref{colfor}) is due to the presence of the additional $\kt$ integration in 
Eq.~(\ref{ktfor}). Nonetheless, $\kt$-factorization can be recast in a form
that is fully consistent with collinear factorization~\cite{CCH,CH}. One first
solves the BFKL equation and factorizes the $\kt$ dependence in Eq.~(\ref{ktfg})
with respect to scale-dependent parton densities. Then, one inserts this 
$\kt$-dependent factor into Eq.~(\ref{ktfor}) and explicitly carries out the 
integration over $\kt$. As a result, one ends up with 
Eqs.~(\ref{colfor},\ref{eveq}) but coefficient functions and splitting functions 
are no longer evaluated in fixed-order perturbation theory. They are indeed
supplemented with the all-order resummation of the logarithmically-enhanced
contributions $\as^n \ln^m x$ at small $x$. $\kt$-factorization is thus a 
powerful tool: the {\em infinite} perturbative summation of the logarithmic 
terms in $C^a(x,\as)$ and $P_{ab}(x,\as)$ is accomplished by a {\em fixed-order}
computation of the off-shell cross section 
${\hat \sigma}(x,Q^2;\kt)$~\cite{CCH,CH} and of
the kernel of the BFKL equation~\cite{FLF,DD,CCq}. 

Note that the coefficient functions and the parton densities on the right-hand
side of Eq.~(\ref{colfor}) are not separately physical observables. One has 
some freedom in redefining them (and, correspondingly, the splitting functions 
in Eq.~(\ref{eveq})), provided that their convolution, i.e. the physical cross
section on the left-hand side of Eq.~(\ref{colfor}), is left unchanged. This 
freedom is called factorization-scheme dependence. The steps leading from the 
$\kt$-factorization formulae (\ref{ktfor},\ref{ktfg}) to resummed splitting 
and coefficient functions can be performed by having full control of the 
factorization-scheme dependence~\cite{CCHLett,CH}. This property is essential 
for the consistency between $\kt$-factorization and collinear factorization.

To the purpose of discussing the main theoretical results of the resummation
programme, it is convenient to introduce
the $N$-moments. For any function $g(x,\dots)$, I define its $N$-moments 
$g_N(\dots)$ in the usual way:
\be
g_{N}(\dots) \equiv \int^{1}_{0} dx \; x^{N-1} \;g(x,\dots) \;\;.
\ee
In particular, the {\em anomalous dimensions} $\ga_{ab, \,N}(\as)$ are related 
to the $N+1$-moments of the Altarelli-Parisi splitting functions, that is,
$\gamma_{ab,N}(\as) \equiv P_{ab, \,N+1}(\as)$.
Note that logarithmic contributions of the type $\ln^{n-1}x$
in $x$-space correspond to multiple poles $(1/N)^{n}$ for $N \to 0$ 
in $N$-space.

As for the parton flavours $a$, it is useful to consider singlet
($a=g,S$) and non-singlet ($a=\{a_{NS}\}$) terms. Thus
${\tilde f}_g$ is the gluon density, and the singlet-quark density is related
to the quark (${\tilde f}_{q_{f}}$) and antiquark 
(${\tilde f}_{{\bar q}_f}$) densities 
as ${\tilde f}_S=\sum_{f} ( {\tilde f}_{q_f}+ {\tilde f}_{{\bar q}_f} ) $.

The first general results following from the $\kt$-factorization theorem regard
the classification of the small-$x$ logarithmic contributions. One can 
show~\cite{CH} that logarithmic terms in the non-singlet sector 
(coefficient functions and splitting functions) are suppressed by (at least) a
power of $x$ with respect to similar terms in the singlet sector. Thus, 
non-singlet corrections are negligible at small $x$ and are not considered in 
the following (they are reviewed elsewhere in these Proceedings~\cite{Blu}).
In the singlet channel one can show that the perturbative expansions of 
the splitting functions and of the coefficients functions for totally 
inclusive cross sections contain at most one power of $\ln x$ (or $1/N$, in $N$
space) for each power of $\as$. Thus, one can set up an improved perturbative 
expansion at small $x$ by systematically resumming the leading logarithmic (LL)
contributions $(\as/N)^n$, the next-to-leading logarithmic (NLL) terms 
$\as(\as/N)^n$ and so forth. 

Among the various anomalous dimensions and coefficients functions, only the
gluon anomalous dimensions $\ga_{gg, \,N}$ and $\ga_{gS, \,N}$ have LL 
contributions. They are given by 
\be\label{andimb}
\!\!\!\!\gamma_{gg, \,N}(\as) = \frac{C_A}{C_F} \;\gamma_{gS, \,N}(\as) 
+ {\cal O}\!\left(\as\left(\frac{\as}{N}\right)^n\right) = 
\gamma_{N}(\as) + 
{\cal O}\!\left(\as\left(\frac{\as}{N}\right)^n\right) , 
\ee
where $\ga_N(\as)$ is the celebrated BFKL anomalous 
dimension~\cite{Jar,CCH,CH}, whose explicit expression will not be reported 
here. 

All the NLL calculations carried out during the last few years 
lead to analytic formulae given in terms of the BFKL anomalous
dimension $\gan$. The NLL contributions to the quark anomalous dimensions
$\ga_{Sg, \,N}$ and $\ga_{SS, \,N}$ have been computed in Ref.~\cite{CH}.
Among the NNL terms in the gluon anomalous dimensions, those proportional to 
$N_f$ have been evaluated recently~\cite{CCq}. The calculation of the remaining
NLL contributions to $\gamma_{gg, \,N}$ and  $\gamma_{gS, \,N}$ is in 
progress~\cite{FLF,DD,CCq}. The processes whose coefficient functions are known
to NLL accuracy are the following: the structure function $F_2$ and the 
longitudinal structure function $F_L$ in DIS~\cite{CH}, the heavy-flavour
photoproduction~\cite{CCH} and hadroproduction~\cite{CCH,CE,LRSS} 
cross sections, the heavy-flavour cross sections via 
intermediate-vector-boson 
exchange~\cite{CC} and the heavy-quark structure functions in 
DIS~\cite{CCH}. Since preliminary experimental results on charm production in 
DIS at HERA have been presented at this Workshop~\cite{Charm}, in the following
Section I review the resummed formulae for the heavy-quark structure functions.

Note that all the NLL results summarized above are factorization-scheme 
dependent~\cite{CCHLett,SDIS}. The factorization scheme that is actually used 
has to be always specified.

\section{Heavy-flavour structure functions in DIS}

The heavy-flavour structure functions $F_2^{Q{\bar Q}}, \,F_L^{Q{\bar Q}}$
are completely analogous to the customary DIS structure functions $F_2, \,F_L$
with the only additional constraint that heavy quarks of mass $M$ are produced
in the final state. The collinear factorization formula can be written as
follows $(i=2,L)$
\be\label{facqq}
F_i^{Q{\bar Q}}(\xi,Q^2;M^2) = \sum_a \int_{\xi}^1 \frac{dz}{z} 
\;C_i^{Q{\bar Q}, \,a}(\xi/z,\as(Q^2);Q^2/M^2) 
\;{\tilde f}_a(z,Q^2) \;\;. 
\ee
Note that in Eq.~(\ref{facqq}) I have defined $F_i^{Q{\bar Q}}$ as function
of $Q^2, M^2$ and the inelasticity variable $\xi$, which is related to the 
customary Bjorken variable $x$ by $\xi=x(1+4M^2/Q^2)$. Since 
$0 \leq \xi \leq 1$, considering the $N$-moments of 
$F_i^{Q{\bar Q}}(\xi,Q^2;M^2)$ with respect to $\xi$, one obtains
\be\label{facqqn}
F_{i, \,N}^{Q{\bar Q}}(Q^2;M^2) = \sum_a  
\;C_{i, \,N}^{Q{\bar Q}, \,a}(\as(Q^2);Q^2/M^2) 
\;{\tilde f}_{a, \,N}(Q^2) \;\;. 
\ee

The coefficient functions $C_i^{Q{\bar Q}, \,a}$ have been computed up to 
one-loop order in Ref.~\cite{QQ}. The corresponding resummed formulae
to NLL accuracy were obtained in Ref.~\cite{CCH}. The gluon coefficient
functions are given as follows
\be\label{cgqq}
\!\!\!\!C_{i, \,N}^{Q{\bar Q}, \,g}\!\!\left(\as;\frac{Q^2}{M^2}\right) 
= R_N \;\frac{Q^2 \;h(\gan)}{16\pi^2 \alpha M^2}
\left(\frac{M^2}{Q^2}\right)^{\gan}
\left(1+\frac{Q^2}{4M^2}\right)^N 
K_N^{(i)}\!\left(\frac{Q^2}{M^2}\right) ,
\ee
where $\gan=\gan(\as)$ is the BFKL anomalous dimension in Eq.~(\ref{andimb})
and $R_N=R_N(\as)$ is a factorization-scheme dependent term (it is the same   
in the \ms\ and DIS schemes: see Eq.~(3.17) of the second
paper in Ref.~\cite{CH}). The functions $h(\ga)$ and $K_N^{(2)}(Q^2/M^2)$ are
respectively given in Eqs.~(3.9) and (4.18) of the second
paper in Ref.~\cite{CCH}. The explicit expression for 
$K_N^{(L)}(Q^2/M^2)$ was not reported in Ref.~\cite{CCH}. It is:
\bea
&&\!\!\!\!\!\!\!\!\!K_N^{(L)}\!\left(\frac{Q^2}{M^2}\right) = 
\left(1+\frac{Q^2}{4M^2}\right)^{-(N+1)} 
\frac{3}{(7-5\gan)(1+2\gan)} \left\{ \left[ 1-\gan + \frac{6M^2}{Q^2} \right]
\right. \nonumber \\
&&\!\!\!\!\!\!\!\!\!\left. 
+ 
\left(1+\frac{Q^2}{4M^2}\right)^{\gan-1}
\left[ \gan(1-\gan)\frac{Q^2}{2M^2} - 2(1-\gan) - \frac{6M^2}{Q^2}
  \right] \right. \\
&&\!\!\!\!\!\!\!\!\!\left. 
\cdot \;F(1-\gan,1/2;3/2;\frac{Q^2}{Q^2+4M^2}) \right\} \;\;,
\nonumber
\eea
where $F(a,b;c;z)$ is the hypergeometric function.
In the \ms\ and DIS schemes the NLL contributions to 
the singlet-quark coefficient functions
are obtained from those in the gluon coefficient functions by the following 
colour-charge relation
\be
\!\!C_{i, \,N}^{Q{\bar Q}, \,S}\!\left(\as;\frac{Q^2}{M^2}\right)  = 
\frac{C_F}{C_A} \left[
C_{i, \,N}^{Q{\bar Q}, \,g}\!\left(\as;\frac{Q^2}{M^2}\right) - \frac{\as}{2\pi} 
C_{i, \,N}^{Q{\bar Q}, \,g \,(0)}\!\left(\frac{Q^2}{M^2}\right) \right] \;,
\ee
where  
the second term in the square bracket is the lowest-order term 
in the perturbative expansion of  $C_{i, \,N}^{Q{\bar Q}, \,g}(\as,Q^2/M^2)$.

\section{Factorization-theorem invariants at small $x$}

So far, the most important phenomenological consequence of small-$x$ 
resummation is its large effect~\cite{EHW,BFres,FRT} on the scaling violation 
of the proton structure function $F_2$ in the HERA kinematic region. To asses
the relevance of this result, further investigations, in particular on the
impact of subleading contributions, are necessary. At the same time,
however, it is worthwhile pointing out that the factorization-scheme 
dependence can play an important role.

The freedom of arbitrarily choosing the factorization scheme is not a particular
feature of small-$x$ dynamics. In general, its
effect amounts to a redefinition of the input parton densities that is 
perturbatively under control. The effect, instead, can be quite large in the
small-$x$ region because each power of $\as$ can be accompanied by an enhancing
logarithmic factor of $\ln 1/x$. 

After having fixed splitting and coefficient functions (using either fixed-order
or resummed perturbation theory), the QCD analysis of the 
HERA data on $F_2$ requires to fit the $x$-dependence of the parton densities at 
some input scale $Q_0$. As a result one may find the dominant small-$x$ 
behaviour ${\tilde f}_g(x,Q_0^2) \sim {\rm const.}$ as well as a steeper 
behaviour ${\tilde f}_g(x,Q_0^2) \sim x^{-\lambda}$. No matter the behaviour one 
finds, one would be
led to conclude that it is due to non-perturbative phenomena. 
Actually, by simple power counting one can argue~\cite{phyad} that
the difference between these two extreme behaviours can consistently be
attributed to different factorization schemes or, equivalently,
to higher-order (i.e. beyond the LL accuracy) logarithmic contributions 
in perturbation theory.
 
This argument may appear very formal (it is indeed provoking, to some 
extent) and with no physical content. Actually, 
the factorization-scheme dependence, rather than an ambiguity in 
higher-order perturbative coefficients, has to be regarded more 
physically as a parametrization of our ignorance in factorizing 
perturbative from non-perturbative physics. Studying the proton structure
function we have at our disposal two quantities, $F_2$ and $dF_2/d\ln Q^2$,
and, besides perturbative QCD, we have to introduce two non-perturbative
distributions, ${\tilde f}_S$ and ${\tilde f}_g$. Roughly speaking, the 
sole proton
structure function does not provide sufficient information to fully disentangle 
perturbative from non-perturbative dynamics at small-$x$. 

A better theoretical control on perturbative physics can be achieved by 
exploiting the very physical content of the collinear-factorization theorem, 
that is, the universality (process independence) of the parton densities. This 
means that the same parton densities and the same perturbative approach have to 
be used to study the small-$x$ behaviour of different physical observables.
Considering more observables, besides $F_2$, one can (over-)constrain the 
definition of the parton densities thus eliminating the factorization-scheme
dependence and emphasizing the perturbative QCD component.

The $\kt$-factorization approach shares these universality features and 
provides resummed perturbative calculations
for many different processes. Thus, one can introduce quantities that are 
invariant with respect to factorization-scheme transformations.
As discussed in Refs.~\cite{CCH,CC}, one 
can consider ratios of hadronic cross sections and properly defined 
$K$-factors (the functions $K_N^{(i)}$ in Eq.~(\ref{cgqq}) are an 
example for that) in which the dependence on the parton densities cancels.
These $K$-factors are factorization-scheme independent by definition.
Analogously, by simply using two hadronic observables 
one can formulate the dynamics of scaling violation enterely in terms of 
perturbative quantities that play the role of physical anomalous 
dimensions~\cite{phyad}.

\subsection{Physical anomalous dimensions}

Considering two different hadronic observables $F_A$ and $F_B$ (for instance,
$F_2$ and $F_L$ or $F_2^{Q{\bar Q}}$), one can write down evolution equations
in the following form
\bea\label{peveq}
\frac{dF_{A} (x,Q^2)}{d\ln Q^{2}} = \int^1_x \frac{dz}{z} 
\!&&\!\!\!\!\!\!\!\!\!\!\!
\left[ \;
\,\G_{AA}(x/z,\as(Q^2)) \;
F_A(z,Q^2) \right. \nonumber \\
&+\!\!&\!\!\left. \G_{AB}(x/z,\as(Q^2)) \;F_B(z,Q^2) 
\;\right] \;\; .
\eea
The parton densities have completely disappeared.
Equation (\ref{peveq}) relates the
scaling violation of two physical observables
to the actual value of the same observables. Thus the
kernels $\G_{AB}(x,\as)$ are {\em physical observables}
as well. Owing to the formal resemblance to the Altarelli-Parisi splitting 
functions in Eq.~(\ref{eveq}), the kernels $\G_{AB}(x,\as)$ can be considered 
as {\em physical splitting functions} and, correspondingly, their moments
$\G_{AB, \,N}(\as)$ are {\em physical anomalous dimensions}.

Each of the physical splitting functions is factorization-scheme invariant
and thus unambiguously computable in perturbation theory. 
In other words, from the
viewpoint of perturbative QCD, each $\G_{AB}(x,\as)$ is completely
analogous to the celebrated ratio 
$R_{e^+e^-} = \sigma(e^+e^- \to {\rm hadrons})/\sigma(e^+e^- \to \mu^+\mu^-)$ in
$e^+e^-$ annihilation.

Although the kernels $\G_{AB}$ can be evaluated without carrying out 
any factorization procedure, they can be related~\cite{phyad} to the customary 
splitting and coefficient functions. Therefore, for any process whose 
coefficient functions are known to one-loop order, one can straightforwardly
obtain the corresponding physical anomalous dimensions to the same accuracy.
Analogously, using the resummed calculations listed in Sect.~\ref{colsec}, one
can derive resummed expressions for the physical anomalous dimensions. In the
case of $\{F_A,F_B\}=\{F_2,F_L\}$, for instance, the anomalous dimension 
$\G_{LL}$ contains LL terms and they are simply given by the BFKL anomalous 
dimension:
\be\label{gllr}
\G_{LL, \,N}(\as) = \ga_N(\as) \;
                    + {\cal O}\left( \as(\as/N)^n \right)  \;\;,
\ee
while $\G_{2L}$ (which drives the scaling violation in the quark channel)
enters only to NLL order and is given by
\be\label{g2lr}
\frac{\as}{2\pi} \;\G_{2L, \,N}(\as) = 
\frac{\as}{2\pi} \;\left[ \frac{1}{1-\ga_N(\as)} + 
\frac{3}{2} \;\ga_N(\as) \right] \;
           + {\cal O}\left( \as^2(\as/N)^n \right) \;\;.
\ee
The expression (\ref{g2lr}) is remarkably simple in comparison with 
those~\cite{CCH,CH,CCq} of other factorization-scheme-dependent quantities 
to NLL accuracy.

In the evolution equation (\ref{peveq}) the small-$x$ perturbative dynamics
is completely controlled by the physical anomalous dimensions.
No subtle theoretical interplay between perturbative logarithms and steepness 
of parton densities takes place. The physical anomalous
dimensions can be evaluated both in fixed-order perturbation theory and in 
resummed perturbation theory. For any given set of observables and kinematic 
region of $x$, one can thus compare the two approaches and study the theoretical
accuracy of the perturbative expansion. Having done that, one can go back to the
customary partonic picture and investigate more safely the 
small-$x$ behaviour of the non-perturbative parton densities. 

A more detailed discussion on physical anomalous dimensions can be found 
elsewhere~\cite{phyad} and some phenomenological studies have already been 
presented at this Workshop~\cite{Thorne}.

\vskip 0.3cm
\noindent{\bf Acknowledgments.}
The results reviewed in this contribution are due to a long and intense
collaboration with Marcello Ciafaloni and Francesco Hautmann.
This research is supported in part by EEC Programme 
{\it Human Capital and Mobility}, Network {\it Physics at High
Energy Colliders}, contract CHRX-CT93-0357 (DG 12 COMA). I am happy to thank
the Local Organizing Committee, and in particular Giulio D'Agostini, for the
organization of an excellent Workshop.

\end{document}